\documentclass[12pt]{article}
\usepackage{graphics}
\usepackage{graphicx,amsmath,amsthm,mathrsfs,amssymb}
\newcommand{\ee}{\end{equation}}
\newcommand{\word}[1]{\,\,\mbox{#1}\,\,}

\newcommand{\beq}{\begin{equation}}
\newcommand{\eeq}[1]{\label{#1}\end{equation}}
\newcommand{\beg}{\begin{equation*}}
\newcommand{\eeg}{\end{equation*}}

\newcommand{\eq}{\!=\!}
\newcommand{\p}{\!+\!}
\newcommand{\m}{\!-\!}

\newcommand{\bsplit}{\begin{split}}
\newcommand{\esplit}{\end{split}}

\usepackage{graphicx}
\usepackage{enumerate}
\usepackage{breqn}

\def\p{{\partial}}

\usepackage{multicol} 
\usepackage{times}
\date{}
\begin{document}


\title{Geometric aspects of Extremal Kerr black hole entropy}
\author{E M Howard \\
~Dept.\ of Physics and Astronomy, Macquarie University, Sydney, Australia.
\\ E-mail: howard@centralcoastlearning.cnri.edu \\ }


\maketitle
\begin{abstract}

Extreme Black holes are an important theoretical laboratory for exploring the nature of entropy. We suggest that this unusual nature of the extremal limit could explain the entropy of extremal Kerr black holes. The time-independence of the extremal black hole, the zero surface gravity, the zero entropy and the absence of a bifurcate Killing horizon are all related properties that define and reduce to one single unique feature of the extremal Kerr spacetime. We suggest the presence of a true geometric discontinuity as the underlying cause of a vanishing entropy.  

\vspace*{5mm} \noindent PACS: 04.62.+v, 04.70.Dy, 04.20.Dw
\end{abstract}

\begin{multicols}{2} 

\section{Introduction}

One of the most remarkable ideas in black hole theory is the 
analogy between the laws of classical black hole mechanics and the laws of thermodynamics. 

Black hole thermodynamics has become an active area of research since
Bekenstein showed that the entropy of a black hole is proportional
to the area of the horizon. It is a well known fact now that that a black hole exhibits 
an unusual similarity to a thermodynamic system. 

Our analysis reveals a purely geometrical disparity between the extreme
and near extreme Kerr geometries, due to the singular nature of extreme regime. 
In other words, the approach to extremality is not continuous.
The nature of the extremal Kerr metric is very different from other stationary solutions.
We focus on relating the entropy of extremal Kerr black holes strictly to their geometric structure. 
Any classical method involving a finite number of steps used for the extremal case leads to subtle inconsistencies like a vanishing entropy and zero surface gravity while the area of the event horizon still remains positive. Using the near-extremal limit to evaluate the black hole entropy leads to major discontinuities. 

Our aim is to understand this discontinuity working solely on pure geometric grounds. 
The discontinuous nature of entropy during the transition from non-extremal to extremal black hole is directly connected with a discontinuous topological nature of the horizon. The entropy of extremal black holes can't be determined as a limit
of the non-extremal case. The geometry of the extreme black holes could shed some light on understanding black hole entropy in general. 
Extremal black holes can't be regarded as limits of non-extremal
black holes due to this discontinuity. We suggest that the reason for this discontinuity is that non-extremal
and extremal black holes are topologically different and the switch from 
one to another can't be done in a continuous manner. In this paper we study various properties of extreme
Kerr black holes to expose the underlying topological nature of the discrepancy between extreme and non-extreme regimes.

\section{No evolution or time reversibility}

It is well known that Einstein's field equations are time-reversal invariant. A maximally extended spacetime includes apart from the black hole solution, its "time-reverse" case. In the non-extreme case, the extended spacetime ($\kappa\ne0$) possesses a bifurcate Killing horizon. In the extreme case (surface gravity $\kappa\eq0$), no distinct time-reverse equivalent exists, the black hole is time-independent everywhere and possesses a single degenerate Killing horizon. The surface gravity $\kappa$ of a black hole cannot be reduced to zero within a finite time. In the extended Schwarzschild spacetime, a white hole region becomes the time-reverse of the black hole region. The future event horizon, which separates two regions, is distinct from the past event horizon. The two regions intersect at the bifurcation two-sphere. The time-reverse of one region yields another region. In the non-extremal Kerr extended case, a new region is the time-reverse of the black hole, generating distinct patches in between the two horizons. The extremal Kerr extended spacetime does not contain such a distinct time-reverse patch but always duplicates of the same patch. There is no distinct time-reverse region and no distinct event horizon. The extremal Kerr black hole has no time-reverse equivalent or, in other words, it is time-independent everywhere. Furthermore, a naked singularity is composed of one single region and has no time-reverse region, has no event horizon and zero entropy. 

In the extremal case, the Killing vector field on the horizon is 
null on a timelike hypersurface intersecting the horizon and it is
spacelike on both sides. 
The event horizon is determined by a Killing vector field whose causal properties 
change from timelike to spacelike across the horizon. This Killing horizon becomes null on a timelike hypersurface surrounding  the horizon. The presence of a Killing vector field which is timelike 
in a region around the event horizon is a very peculiar and puzzling feature.
The horizon Killing field is spacelike except at the horizon itself. 

In Boyer-Lindquist coordinates, the Kerr metric is given by

\begin{equation}
\begin{split}
 ds^{2} = -\left(1-\frac{2Mr}{{\rho}^2}\right)dt^2 - \frac{4Mar{\sin^2{\theta}}}{\rho^2}dtd\phi  \\
+ \frac{\Sigma}{\rho^2}{{\sin^2{\theta}}}d\phi^2 +  
\frac{\rho^2}{\Delta}dr^2 + {\rho^2}d\theta^2.  \hspace*{+2.4cm}
\end{split}
\end{equation} 

where
\begin{eqnarray}
 \rho^2 &=& r^2 + a^2{\cos}^2\theta. \\
\Delta &=& r^2 + a^2 - 2Mr.\\
\Sigma &=& \left(r^2 + a^2\right)^2 - a^2\Delta{\sin}^2\theta. 
\end{eqnarray}
The horizons are situated at $\Delta = 0 $, i.e at,
\begin{equation}
 r_{\pm} = M \pm \sqrt{M^2 - a^2}.
\end{equation}

\noindent The Kerr spacetime gets divided into three regions: 
\begin{eqnarray}
\mbox{I}: \ r_+ < r < \infty \ , \\
\mbox{II}: \ r_- < r < r_+ \ , \\
\mbox{III}: - \infty < r < r_- \  . 
\end{eqnarray}

\noindent Region I is the exterior, region II lies between 
the two Killing horizons at $\Delta = 0$, and region III is an
asymptotic region. The mass of the black hole is $m$, its angular momentum $J = am$, 
and its event horizon happens at $r = r_+$. The extremal case is obtained by setting 
$a = m$, in which case there is no region II since the two horizons coincide. The 
$a < m$ case describes the generic black hole, while $a > m$ describes a naked singularity.

The solutions for $\Delta\eq 0$ generate the inner and outer horizons:

 $r_{\pm}=m \pm\sqrt{m^2-a^2}$.
 
The metric has two Killing vectors, $K^{\mu}\eq (\partial_t)^{\mu}\eq (1,0,0,0)$ and  $R^{\mu}\eq (\partial_{\phi})^{\mu}\eq (0,0,0,1)$. We can construct a Killing vector $\chi^{\mu}$ as a linear combination of two vectors: $\chi^{\mu}\eq  K^{\mu} +\omega_0\,R^{\mu}$ with $\omega_0$ being a constant. 

Both $\rho^2$ and $\Sigma$ are positive but on the ring singularity they vanish. 

In the non-extremal case, $\Delta$=$(r-r_{\p})(r-r_{\m})$.
$\Delta$ is positive for $r\!>\! r_{\p}$ or $r\!<\! r_{\m}$ and zero for $r\eq r_{\p}$ and $r\eq r_{\m}$. 
But $\Delta$ is negative in the region $\Re$ representing $r_{\m}\!<\!r\!<\!r_{\p}\,\,$, always remaining between the horizons. Since any linear combination of the two Killing vectors remains spacelike, $\Re$ is nonstationary.  

In contrast, the extreme case ($a\eq m$) has $\Delta$=$(r-a)^2$, positive everywhere except at the event horizon ($r \eq a$) where it is zero. If $\omega_0$=$\omega$ at $r\eq r_0$, we obtain the angular velocity of a zero angular momentum observer (ZAMO). A ZAMO observer exists at every point in the spacetime. The Killing vector $\chi^{\mu}$ is timelike or null. At each point, one can have a different timelike Killing vector (except at the horizon where it vanishes). There is no point where all Killing vectors are spacelike and no region is non-stationary. A general Killing vector field of the generic Kerr solution
is given by the linear combination 
\begin{equation} \xi = \partial_t + K\partial_\phi \ , \end{equation}

\noindent where $K$ is a constant. Its norm is 
\begin{eqnarray} ||\partial_t + K\partial_\phi ||^2 = \frac{1}{\rho^2}[ 
-\Delta + \sin^2{\theta}(a^2-    \nonumber  \\
4marK + AK^2)] \  \hspace*{+3.5cm}
\end{eqnarray}

\noindent This can vanish for all $\theta$ only if $\Delta = 0$, that is 
on one of the horizons. It vanishes on the outer horizon $r = r_+$ if and only if 

\begin{equation} K = \frac{a}{2mr_+}. \  \end{equation}

\noindent The linear combination defines the horizon Killing 
vector field $\xi_{\rm hor}$. We have

\begin{equation} || \xi_{\rm horizon}||^2 = - \frac{r-r_+}{4m^2\rho^2r_+^2}f(a,r,\theta) \ , 
\label{norm} \end{equation}

\noindent where

\begin{eqnarray} f(a, r,\theta) = - 4m^2a^2\sin^2{\theta}(r-r_+) + \nonumber \\
+( 4m^2r_+^2 - a^2(r^2+2mr+a^2)\sin^2{\theta} + \nonumber \\
a^4\sin^4{\theta})(r-r_-)  \ . 
 \hspace*{+3.5cm}
\end{eqnarray} 

In $3+1$ decomposition, the Kerr metric becomes:


\begin{dmath}
ds^2 = N^2\,dt^2 + h_{ab}\,(dy^a +N^a \,dt)(dy^b +  \\  
N^b\,dt) = N^2\,dt^2 +h_{\phi\phi}(d\phi +N^{\phi}\,dt)(d\phi +  \\ 
N^{\phi}\,dt) +h_{rr}\,dr^2 +h_{\theta\theta}\,d\theta^2 \\ 
=\dfrac{\rho^2 \,\Delta}{\Sigma}\, dt^2 - \dfrac{\Sigma\sin^2\theta}{\rho^2} \,(d\phi-\omega\,dt)^2 -  
\dfrac{\rho^2}{\Delta}\,dr^2 -\rho^2\,d\theta^2  \\
\end{dmath}
where
\begin{eqnarray}
N^2 = \dfrac{\rho^2 \,\Delta}{\Sigma}, h_{\phi\phi}=  
- \dfrac{\Sigma\sin^2\theta}{\rho^2}, h_{rr} \eq -\dfrac{\rho^2}{\Delta}, \nonumber \\
h_{\theta\theta}\eq -\rho^2 \word{and} N^{\phi} = -\,\omega\,.  \nonumber \\
\hspace*{+5cm}
\end{eqnarray}

where we introduced the lapse function $N$, the shift vector $N^a$ and the induced metric $h_{ab}$.

Depending on $\Delta$, the solution depicts a black hole or a naked singularity. The lapse function is never negative. Since $\rho^2$, $\Sigma$ and $\Delta$ depend only on $r$ and $\theta$, the lapse function $N$ also depends only on $r$ and $\theta$ and is time-independent. 
The induced $h_{ab}$ and the momentum conjugate $P^{ab}$ are time-independent everywhere. Thus, there is no time evolution of the phase space $[h_{ab},P^{ab}]$. 
The coefficients $h_{\phi\phi}, h_{rr}$ and $h_{\theta\theta}$ can't be positive and they are independent of time. They all depend on $r$ and $\theta$: $\dot{h}_{ab}$ is zero everywhere. $N^{\phi}$ is also time-independent and it depends only on $r$ and $\theta$. The momentum conjugate to $h_{ab}$, $P^{ab}$ is also time independent. The extrinsic curvature $K_{ab}$ is:
\beq
\begin{split}
&K_{ab}\equiv\dfrac{1}{2N} (\dot{h}_{ab}-\nabla_b\, N_a-\nabla_a\, N_b)\,. \end{split} \eeq{gfd}
with the components:\\
\beq
\begin{split}
&K_{\phi\theta}=K_{\theta\phi}=\dfrac{1}{2N}(-\partial_{\theta} N_{\phi} +\dfrac{1}{2} \,N^{\phi}\partial_{\theta} h_{\phi\phi})
\end{split}
\eeq{gfd}
and
\beq
\begin{split}
K_{\phi\,r}=K_{r\,\phi}=\dfrac{1}{2N}(-\partial_{r} N_{\phi} +\dfrac{1}{2} \,N^{\phi}\partial_{r}h_{\phi\phi})\\
\end{split}
\eeq{gfd}
where
\beq
\begin{split}
N_{\phi}=h_{\phi\phi}\,N^{\phi} = \dfrac{\Sigma\,\omega\,\sin^2\theta}{\rho^2} 
\end{split}
\eeq{gfd}
and
\beq
\begin{split}
K\equiv h^{ab}\,K_{ab} =0\,.
\end{split}
\eeq{gfd}
The momentum conjugate is:
\beq
\begin{split}
&P^{ab}=\dfrac{\sqrt{-h}}{16\pi}\,(K^{ab} - K\,h^{ab})= \dfrac{\sqrt{-h}}{16\pi}\,K^{ab}
\end{split}
\eeq{gfd}
with the components:\\
\beq
\begin{split}
&P^{\phi\theta}=P^{\theta\phi}= \dfrac{\sqrt{-h}}{16\pi}\,K^{\phi\theta} 
\end{split}
\eeq{gfd}
and
\beq
\begin{split}
P^{\phi\,r}=P^{r\,\phi}= \dfrac{\sqrt{-h}}{16\pi}\,K^{\phi\,r}\\
\end{split}
\eeq{pab2}
where
\beq
\begin{split}
K^{\phi\theta} = \dfrac{K_{\phi\theta}}{h_{\phi\phi}h_{\theta\theta}} \word{and} K^{\phi\,r} = \dfrac{K_{\phi\,r}}{h_{\phi\phi}\,h_{r\,r}} \,.
\end{split}
\eeq{pab2}
Obviously $P^{ab}$ is only dependent on $r$ and $\theta$ and consequently it is independent of time. There is only one single classical microstate $[h_{ab}(r,\theta), P^{ab}(r,\theta)]$ available. 

In the non-extremal case, the region $\Re$ is non-stationary and the general $3+1$ decomposition in this region has $N^2$ positive. However, all $\Re$ components are time dependent. The phase space in $\Re$ is time dependent. The entropy is not vanishing, since there is more than one classical microstate hidden from the outside observer.

\section{No thermality}

A non-extremal spacetime with an outer and an inner horizon becomes
extremal as the outer horizon approaches the inner one. 
The two horizons are in equilibrium at two different temperatures and
the temerature of the outer horizon approaches zero. Consequently,
no thermality is observed at the outer horizon.

In the extremal regime, the flux is the same
both outside and inside the horizon and approaches zero on the horizon, 
with a vanishing temperature in the extremal case. There is a finite
discontinuity of flux in the extremal limit of the non-extremal
regime. There is no physically acceptable smooth transition from 
the extremal regime as a limiting case of the non-extremal one and
the pure extremal case.

To understand the thermal nature of an extreme black hold, let us firstly find the probability flux across the horizon. To achieve this, we write the Kerr metric in Kruskal coordinates and set $a = m$ for the extreme regime:
\beq
\begin{split}
 ds^2 = - \left(1-\frac{2mr}{{\rho}^2}\right)dt^2 - \frac{4m^2r{\sin^2{\theta}}}{\rho^2}dtd\phi + \\
 \frac{\Sigma}{\rho^2}{{\sin^2{\theta}}}d\phi^2 + 
\frac{\rho^2}{\Delta}dr^2 + {\rho^2}d\theta^2.
\end{split}
\eeq{pab2}
with
\begin{eqnarray}
 \rho^2 &=& r^2 + m^2{\cos}^2\theta. \\
\Delta &=& \left(r-m\right)^2 \\
\Sigma &=& \left(r^2 + m^2\right)^2 - m^2\Delta{\sin}^2\theta 
\end{eqnarray}
The horizon is located at $r = m$. Introducing the null coordinates $u = t - r_*, v = t + r_*$ with

\begin{eqnarray}
u &=& -M\cot U.\\
v &=& - M\tan V.
\end{eqnarray}

and 
\begin{eqnarray}
r_* = \int\frac{r^2 + m^2}{\left(r-m\right)^2}dr =   \nonumber \\
r + 2m\ln|r - m| - \frac{2m^2}{r-m}.\\ \hspace*{+5cm} \nonumber 
\end{eqnarray}
\begin{equation}
\tilde\phi = \phi - \frac{1}{2m}t = \phi - {\Omega}t
\end{equation}
the surface $r = m$ appears at $v-u = -\infty$. The Kruskal coordinates $U,V$ are:

The future horizon is located at: \\
$U = 0, V < \frac{\pi}{2}$.

The Killing vector $\partial_t + \Omega\partial_{\phi}$ becomes
\begin{eqnarray}
 \partial_t + \Omega\partial_{\phi} &=& \partial_u + \partial_v + \frac{\partial\tilde\phi}{\partial t}\partial_{\tilde\phi} + \Omega\partial_{\tilde\phi} \nonumber \\
    &=&\frac{1}{m}\left[\sin^2{U}\partial_U - \cos^2{V}\partial_V\right]. \nonumber \\
\end{eqnarray}

To find the outgoing probability flux across the horizon, we find:
\beq
\begin{split}
\frac{j^{out}}{\overline\Psi_{\omega}(U)\Psi_{\omega}(U)} =   \\
-\frac{iU^2}{m}\partial_U[\ln \overline\Psi_{\omega}(U) - \ln \Psi_{\omega}(U)] =  \\
= \omega U^2[\frac{1}{(U +i\epsilon)^2} + \frac{1}{(U-i\epsilon)^2}]
\end{split}
\eeq{pab2}

We have $\left(U\pm i\epsilon\right)^{-2} = PV(1/U^2) \pm i\pi\delta^\prime(U)$ and $U^2\delta^\prime(U) = -2U\delta(U) = 0$.
Taking logarithm of both sides we get the flux:
\begin{equation}
 \ln j^{out} = \ln |N_\omega|^2 - 2\pi\omega m\delta(U).
\end{equation}
For any $N\neq 0$, $j^{out} = |N_\omega|^2$, no thermality is observed. The horizon behaves like a transparent membrane. The flux at $U=0$, is obtained by regularizing the delta-function. Using the limit $\epsilon \to 0$, we obtain:
\begin{equation}
 j^{out} = |N_\omega|^2\exp\left(-2m\omega\frac{\epsilon}{U^2 + \epsilon^2}\right).
\end{equation}
The flux $j^{out} = |N_\omega|^2\exp\left(-2m\omega/\epsilon\right)\to 0 $ when $\epsilon\to 0$. A finite discontinuity of flux and no thermality is present at the horizon. This discontinuity is not coordinate related. For an extremal black hole, the proper radial distance from the horizon to any point close to the horizon outside or inside, is infinite. It is impossible for any incident particle state to cross the horizon. An extremal black hole cannot absorb or emit particle states. The coordinates $U_+,V_+$ remove the coordinate singularity at the horizon. However $U_+$ fails to be a proper distance. The discontinuity in $U_+$ is not physical. The flux needs an infinite proper time to become zero at the horizon when it is incident either from inside or outside. When equated to the Boltzmann factor, it implies an infinite $\beta$. This is equivalent to setting a zero temperature for the black hole. 
In the extremal limit, when $r_+ \to r_-$, the effective temperature is zero and the emission probability is also zero. The flux is the same both outside and inside the horizon and zero at the horizon. This means a vanishing temperature, therefore a finite discontinuity in flux.

\section{Vanishing Entropy}
\label{ent}

An interesting question appears. Where should we calculate the entropy: on (or nearby) the horizon or within the black hole (disc) itself?
Is the entropy created immediately after the gravitational collapse, or later during the black hole
evolution? The extremal entropy is independent of the black hole evolution and its internal configuration.
Can such a function be purely derived from geometric or topological considerations?
The third law of thermodynamics states that the surface gravity vanishing limit cannot be reached within a finite time. Cosmic censorship Conjecture forbids reducing the surface gravity to zero. It is impossible through a finite number of physical processes to reach the
zero limit surface gravity. However, the extremal Kerr black hole has a vanishing surface gravity therefore zero temperature. 

Statistical mechanics describes the entropy of a system by the natural
logarithm of internal states count:
S = lnQ.
A microstate Q­ is a function of the system's macrostate. The entropy
is a function of these variables. An interesting
relation between (macroscopic) entropy and statistical thermodynamics (number of microscopic states) becomes evident.
Where there is only one microstate, Q = 1 and the entropy is zero, no disorder is present.
One single state corresponds to the extremal regime. 
Extremal black holes can't be viewed as limits of non-extremal regime
because of this discontinuity. 
Our assessment is that the non-extremal and
extremal regimes are topologically different and the discontinuity itself can be explained on geometric grounds. 

An infinitesimal perturbation in mass, spin and horizon area can be written in Kerr metric:

\begin{equation}
dm~=~\frac{k}{8\pi} dA + \Omega_J da,
\label{bhtd}
\end{equation}
where $m$ and $a$ are the mass and the spin of the black hole.
The horizons form at $r_\pm = m \pm \sqrt{m^2 -
a^2}$. The surface gravity $\kappa$ is 
$(r_+-r_-)/2r_+^2$. The Hawking temperature of
the black hole is $T_H=\kappa/2\pi$. If we compare it with the first law of
thermodynamics,
\begin{equation}
dE~=~ TdS - PdV~,
\label{td}
\end{equation}
and replace the term $PdV$ by $-\Omega_J da$, we find that the entropy $S$
must be identified with $A/4$ up to an affine
constant, which we set to zero by introducing the limit $S_{BH}
\rightarrow 0$ when $m \rightarrow 0$.

The entropy becomes:
\begin{equation}
S_{BH}~=~ \frac{A}{4G}~,
\end{equation}
where $S_{BH}$ is the Bekenstein-Hawking entropy, $G$ is Newton's constant and $A$ is
the event horizon area. 

We see that in extremal regime, $T_H=0$ because $r_+=r_-$. The analogy between the equations (\ref{bhtd}) and (\ref{td}) can't be done anymore. We also have
\begin{equation}
T_H^{-1}~=~\left(\frac{\partial S_{BH}}{\partial m}\right)_a~,
\end{equation}
where the right hand side diverges at the limit $T_H=0$. 
The entropy can be now described a function of $m$ with a singularity at the
extremal limit $m=a$.

Let us assume the entropy being an arbitrary function of the area:
\begin{equation}
S_{BH}~=~f(A)~,
\end{equation}
with
\begin{equation}
\Delta S_{BH}~=~\frac {d~f}{d A} \Delta A~,
\label{sbh}
\end{equation}
where $\Delta S_{BH}$ and $\Delta A$ are the
change in entropy and area of the black
hole when the internal configuration is changed (for example, when a particle falls into it). After the particle crosses
the horizon, there is no information about
its state or status. We therefore assume that
it is equally probable for it to exist
or not. The minimum entropy change can be written
\begin{equation}
(\Delta S)_{min}~=~\sum_n~p_n \ln p_n~=~ \ln2~,
\end{equation}
where the summation over $n$ represents all possible
states of the particle. 
The proper radius 
$b$ for an incident particle with mass $\mu$ and
center of mass at $r_+ + \delta$ is $\int_{r_+}^{r_+ + \delta} \sqrt {g_{rr}} dr$. The
minimum change of black hole area
can be written
\begin{eqnarray}
(\Delta A)_{min}~=~ 2 \mu b~, ~~~~ {\mbox with }\nonumber\\
b~=~2 \delta^{1/2} \frac {r_+}{\sqrt {r_+-r_-}}~.
\end{eqnarray}
and
$g_{rr}~=~ (r-r_-)(r-r_+)/r^2$, 
under the non-extremal condition $(r_+ - r_-\gg \delta)$. 
However, in the extremal limit $(r_+\rightarrow r_-)$, the radius b~ becomes:
\begin{equation}
b~=~\delta~+~r_+\ln(r-r_+)|_{r_+}^{r_+ + \delta}~,
\label{bmin}
\end{equation}
diverging for any $\delta >0$.
For any small finite $\delta$, the proper
radius of the particle becomes infinite. Therefore we need to have $\delta~=~0$. For
$b~=~0$(point particle), we have $(\Delta A)_{min}~=~0$.
The change in entropy requires that
\begin{equation}
\left (\frac {\partial f}{\partial A} \right )_{r_+=r_-} \longrightarrow \infty~,
\end{equation}
with a discontinuity in the
extremal limit.

A semi-classical picture evaluates the gravitational path integral by employing the euclideanized black hole geometry.
If we consider the euclideanised case, the metric in $n$ dimensions near the
horizon is
\begin{equation}
ds^2~=~ N^2 d\tau^2 + N^{-2} dr^2 + r^2~d \Omega_{n-2}^2~.
\end{equation}
and the proper angle $\theta $
in the $r- \tau$ plane near the horizon becomes
\beq
\begin{split}
\theta~=~ 
\frac{\int_{t_1}^{t_2} {\sqrt {g_{\tau \tau}}~d\tau}}
{\int_{r_+}^{r} \sqrt{g_{rr}}~dr}~=  \\
=~ (N N')|_{r_+} (t_2-t_1)  \\
\end{split}
\eeq{pab2}
where $N'$ is the N differentiated with respect to $r$.
$N$ depends on the proper angle $\theta$:
\begin{equation}
(t_2-t_1) N^2~=~2\theta~(r-r_+) + O [(r-r_+)^2]~.
\end{equation}
In two dimensions, the metric near the horizon reduces to:
\begin{equation}
ds^2~=~d\rho^2 + \rho^2 d\theta^2~,
\end{equation}
with $\rho \equiv  \sqrt{2 (r-r_+)/NN'}.$ 
To avoid a conical singularity at the horizon, the period of 
$\theta$ is identified with $2\pi$,  which corresponds to the
topology of a {\it disc} with zero deficit angle 
in the $r-\tau$ plane. This can always be done for non-extremal
black holes, as $(NN')|_{r_+}$ is non-zero.
In the extremal case, the proper radius diverges and the proper angle tends to zero. 
The disk topology is replaced with an annulus one. The topology of the transverse section in
either case is $S^{n-2}$. For a wrong periodicity, the geometry would have a singularity at the origin linked to the excess angle, like a cone structure that we may create out of a sheet of paper.
The conical excess angle becomes $2\pi$ and the topology is that of an annulus. The topology of the transverse section in
either case is $S^{n-2}$. An interesting feature of the solution is that the interior of the black hole is completely absent in the euclidean case. To know what is happening inside the black hole, the euclideanized spacetime is continued to an imaginary value of the radial coordinate near the horizon.

The Euclidean action leads to the entropy
\begin{equation}
S = \left( \beta \frac{d}{d\beta} - 1 \right) I_E .
\end{equation}

The euclideanized version imposes $t_Euclid = i t$ in the metric, with $r > r_+$ leading to a $R^2 \times S^{2}$ manifold topology. The polar coordinates $\{r, t_Euclid \}$ are defined on the $R^2$ with origin $r_+$ and periodicity $\beta$ of the Euclidean time angle $t_Euclid$. An interesting fact is that the Lorentzian manifold $r < r_+$ regions can't be included in the Euclidean solution. 
The boundary contributions from the vicinity of the origin for the non-extremal regime are independent of $\beta$ and the canonical action is proportional to $\beta$, leading to Bekenstein-Hawking entropy $S = A / 4 G$. The $S^2$ contribution does not degenerate at the origin because $r = r_+$. 

In the extremal regime, $r_+$ is infinitely far away from any point outside the horizon and consequently this point must be removed from the whole Euclidean manifold. The extremal black hole horizon is an infinite proper distance from any
stationary observer. There is an absence of a conical singularity at the origin of the manifold. The topology of the Euclidean extremal solution becomes $R \times S^1 \times S^2$. The periodicity of the the Euclidean time is not fixed. Because the origin doesn't exist any more, for any periodicity of the Euclidean time, no conical singularity is formed. The origin is not part of the manifold and the contribution from the vicinity of the origin vanishes. Consequently the entropy is zero. 

We believe that this topology can explain what is actually happening with the entropy. If we consider the black hole as a
microcanonical ensemble, in a
Hamiltonian formulation, the action $I$ is proportional
to the entropy. A dimensional continuation of  Gauss-Bonnet
theorem to $n$ dimensions gives us the action: 
\begin{equation}
I~\propto~ \chi ~ A_{n-2}
\label{act}
\end{equation}
with $\chi$ the Euler characteristic of the euclideanised
$r- \tau$ plane and $A_{n-2}$ the area of the transverse
$S^{n-2}$. 
The black hole entropy becomes
\begin{equation}
S~=~\frac{ \chi A}{4G}~.
\end{equation}
In the non-extremal regime, $\chi =1$ corresponding to a disc and leading to the regular area law.
In the extremal case, we have $\chi=0$ (annulus), implying a
vanishing entropy. 

\section{Discussion}

Cosmic Censorship tries to solve the following problem: can a singularity exist in the absence of a horizon (naked singularity, not a topic of this paper)? Similarly, the existence of extreme black holes relies upon answering the question: can a spinning singularity exist without two separate horizons? This is an interesting problem.
Extremal Kerr black holes are stationary black holes whose inner and outer horizons coincide. 
An extremal black hole has $\kappa \eq 0$ and no bifurcate Killing horizon. Also, the past and future horizons never intersect.
There is no physical process that can make an extremal black hole out of a non-extremal black hole. 
A near-extremal black hole has a sort of potential barrier close to the horizon that prevents to reach extremal regime. These type of black holes represent the asymptotic limit of physical black holes. 
Extremal black holes never behave as thermal objects. Their temperature is always undefined and their emission spectrum is non-planckian. Their non-thermal nature is a consequence of the geometric nature of the horizon.
All characteristics of the stress-energy tensor are different from
those in the non-extremal regime. A final thermal macrostate can't be achieved in a smooth continuous
way without violating the energy conditions. The extremal Kerr black hole cannot be produced through a process involving any finite number of steps without violating the weak energy condition. Consequently an extremal black hole cannot be produced through a finite number of processes or through standard gravitational collapse. 

However, black hole entropy can be also defined as a measure of the observer's accessibility to information about the internal configuration hidden behind the event horizon \cite{Bekenstein}. This internal configuration can be depicted as the sum of points in the phase space, defined by a number of classical microstates $[h_{ab},P^{ab}]$. We can say that extremal Kerr black hole has zero entropy because it possesses a single classical microstate. Extremal black holes have zero temperature, surface gravity $\kappa\eq 0$ and zero entropy therefore they obey the strong version of the third law of thermodynamics. 

We could say that an extremal Kerr black hole has zero entropy only because it has one single classical microstate. There is no continuous set of classical states and no time evolution. All metric components are time independent. Any observer should have complete access to the unique classical state found within the region beyond the event horizon.   

A regular black hole has non-zero entropy because an event horizon hides its internal configuration and there are more than one internal states within its configuration. An extremal black hole has an event horizon but because the phase space is time independent, it doesn't hide more than one internal configuration. The extremal solution doesn't have a time-reverse equivalent or a bifurcate Killing horizon. The condition for a non-zero entropy is the existence of a bifurcate Killing horizon in the extended spacetime. Wald's general formula for the entropy of a black hole should be calculated at the bifurcation two-sphere not on the event horizon. The entropy is calculated as the integral of a geometric quantity over the spacelike cross section of the event horizon. The generated entropy at the event horizon represents the Noether charge of the Killing isometry that generates the event horizon itself. This geometric origin of the entropy suggest a deeper connection between the gravitational entropy, the topological structure of the spacetime and the nature of gravity. We suggest a purely local and geometrical character of the entropy. The different nature of the event horizon in the regular and extreme case requires different calculations of the entropy. The thermodynamic features are consequences of the topological structure of the space-time. The geometric nature of the boundary of the manifold determines the character of the entropy of the black hole. In the non-extremal regime, the proper distance and the coordinate distance between the inner and outer horizons are finite. In the extremal case, the proper distance between the event and Cauchy horizons becomes infinite, even though the coordinate distance vanishes. All these peculiar features are connected to a major property of extremal geometry: the absence of outer trapped surfaces within the horizon, which will be the subject of further work.
Wald also assumed a local, geometrical character of the concept
of black hole entropy, when he calculated the entropy at the bifurcation two-sphere. Entropy is depicted as a local geometrical quantity
integrated over a spacelike region of the horizon.

For a regular black hole, the zeroth law imposes that the
horizon of a black hole must be bifurcate and the surface gravity must be must constant and non-vanishing.
In the case of a degenerate Killing horizon there is no such bifurcation between the horizons. The area between horizons is completely absent.

\section{Conclusion}

The primary condition for a non-zero entropy is the existence of a bifurcate Killing horizon. This single criteria is enough. We analyzed a few features of the Kerr black hole that distinguish the extremal regime from the near-extremal one. From a thermodynamic point of view, there is a discontinuous nature of the entropy between the non-extremal and extremal cases, as the entropy of extremal regime is not the limit of the non-extremal one. While dual and string theory dual microstate counting predict non-vanishing entropy solutions for extreme regime, we suggested that the unusual nature encountered in the extremal limit using semi-classical methods represents a genuine relevant topological discontinuity and therefore the origin of a vanishing entropy. The entropy is zero, in agreement with semi-classical solutions, due to a degeneracy of the horizon geometry. The spacetime topology plays an essential role in the explanation
of intrinsic thermodynamics of the extreme black hole solution.
We conclude that the topology itself of the extreme black hole is enough to explain entropy in this regime. Moreover, the study of extreme black holes could play a crucial role in understanding gravitational entropy in general.

\setcounter{equation}{0} \label{sec7}

\vglue .5cm

\bibliography{apssamp}

\end{multicols} 
\end{document}